# Band-Edge Orbital Engineering of Perovskite Semiconductors for Optoelectronic Applications

Gang Tang,[a, b] Philippe Ghosez,[b] Jiawang Hong [a, *]

[a] School of Aerospace Engineering, Beijing Institute of Technology, Beijing, 100081, China

[b] Theoretical Materials Physics, Q-MAT, CESAM, University of Liège, B-4000 Liège, Belgium

**Corresponding Author**

*E-mail: hongjw@bit.edu.cn.

**ABSTRACT**

Lead (Pb) halide perovskites have achieved great success in recent years due to their excellent optoelectronic properties, which is largely attributed to the lone-pair $s$ orbital-derived antibonding states at the valence band edge. Guided by the key band-edge orbital character, a series of n$s^2$-containing (i.e., $Sn^{2+}$, $Sb^{3+}$, $Bi^{3+}$) Pb-free perovskite alternatives have been explored as potential photovoltaic candidates. On the other hand, based on the band-edge orbital components (i.e., $M^{2+}$ $s$ and $p$/$X^-$ $p$ orbitals), a series of strategies have been proposed to optimize their optoelectronic properties by modifying the atomic orbitals and orbital interactions. Therefore, understanding the band-edge electronic features from the recently reported halide perovskites is essential for future material design and device optimization. Here, this Perspective first attempts to establish the band-edge orbital-property relationship using a chemically intuitive approach, and then rationalizes their superior properties and understands the trends in electronic properties. We hope that this Perspective will provide atomic-level guidance and insights toward the rational design of perovskite semiconductors with outstanding optoelectronic properties.

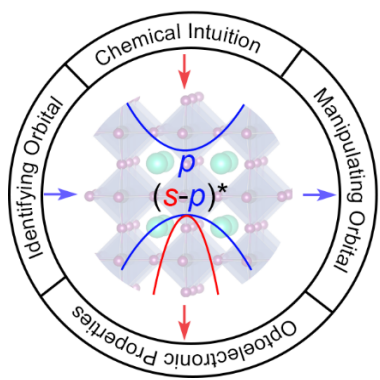

# 1. Introduction

Metal halide perovskites with a general formula $AMX_3$ (A = Cs, $CH_3NH_3$, $CH(NH_2)_2$; M = Pb, Sn; X = I, Br, Cl) have received intensive attention due to their excellent electronic and optical properties for perovskite-based optoelectronic applications.[1-3] Previous theoretical studies[4-6] have revealed that high-symmetry perovskite structure plays a critical role in establishing superior optoelectronic properties. Therefore, in the past few years, exploring the structure-property relationships has become an important guidance for experimental efforts in seeking high-performance halide perovskites.[7-9] For examples, from three- (3D) to two- (2D) to one-dimensional (1D) structures, perovskite semiconductors usually show increased band gap, exciton binding energy, and carrier effective mass, as well as reduced band-edge dispersion and band width.[10-12] Along with the reduction of structural dimensionality, the connectivity of $[MX_6]$ octahedra in $AMX_3$ halide perovskites has also changed significantly.[13] Generally, increasing the octahedral connectivity from corner-, to edge-, to face-sharing causes a significant increase in the band gap and reduction in the band-edge dispersion.[12-14] The combination of high structural dimensionality and corner-sharing connectivity has become a quite important structural descriptor for searching for high-performance photovoltaic materials. However, the recently synthesized halide double perovskites $A_2M(I)M(III)X_6$ (A = monovalent cation, M(I) = Na, Ag, Au, Tl; M(III) = Sb, Bi, Au, In, Tl; X = halogen)[3, 15-20] present a challenge to the structure-property relationship. They usually have a 3D structure similar to $AMX_3$, but most of them have wide indirect band gaps and large carrier effective masses.[21-23] At present, the reported highest power conversion efficiency (PCE) of $Cs_2AgBiBr_6$-based perovskite solar cells (2.79%)[24] is much lower than that of perovskite solar cells based on $CH_3NH_3PbI_3$ or $CH(NH_2)_2PbI_3$ (> 20%)[1-2]. Therefore, in addition to the crystal structure factor, the microscopic electronic structure features must be uncovered to further understand the photovoltaic performance of halide perovskites.

Electronic structure calculations have revealed that the excellent optoelectronic properties of Pb halide perovskites are largely attributed to the lone-pair $s$ orbital-derived antibonding states at the valence band edge.[4, 25-28] Guided by the key orbital character, a series of $n$s$^2$-containing (e.g., Sn$^{2+}$, Bi$^{3+}$, Sb$^{3+}$) halide perovskites have been widely explored as promising photovoltaic materials.[3, 20, 29] Further, Xiao et al.[11] proposed the electronic dimensionality concept, which describes the connectivity of the atomic orbitals that comprise the lower conduction band and the upper valence band, to account for the reported trends in photovoltaic performance of absorbers. Compared with structural dimensionality, this concept gives more reasonable explanation for reported properties such as band gaps, carrier effective masses, defect levels of halide perovskites and nonperovskites. For example, the unfavorable optoelectronic properties of most 3D double perovskites (e.g., $Cs_2AgBiBr_6$) can be well understood from their low electronic dimensionality. It is because that both the valence band maximum (VBM, deriving from Ag 4$d$ and Br 5$p$ orbitals) and conduction band minimum (CBM, deriving from Bi 6$p$ orbitals) of $Cs_2AgBiBr_6$ cannot connect 3-dimensionally due to [AgBr$_6$] octahedra isolated by the adjacent [BiBr$_6$] octahedra.[11, 22] Therefore, the electronic features at the band edges play a critical role in determining the photovoltaic performance of perovskite semiconductors.

Inspired by previous orbital engineering works in layered thermoelectric materials,[30] topological insulators,[31] and perovskite oxide superlattices,[32] we recently introduced the orbital engineering concept to focus on exploring the new orbital-property relationship in halide perovskites.[19, 33] The realization of orbital engineering in halide perovskites can be summarized as two aspects including: (i) identifying the key orbital hybridization character at the band edges to screen and design novel functional materials, and (ii) manipulating the orbital components of band edges to tune and optimize material properties. Regarding the first aspect, a typical example is the use of lone-pair *s* orbital-derived antibonding states at the valence band edge as a screening rule to search for promising optoelectronic materials.[3-4] Another example is the rational control the *d-p* orbital hybridization to achieve flat conduction band and flat valence band simultaneously in a 3D halide perovskite.[19] Regarding the second aspect, adjusting the band-edge orbital components (e.g., $M^{2+}$ *s* and *p*/$X^-$ *p* orbitals) is an important strategy to achieve superior optoelectronic properties in halide perovskites. Common methods including compositional engineering,[34] dimensionality engineering,[10] strain engineering,[35] etc. have been widely reported in perovskite solar cells. Here, this Perspective employ a simplified bonding model to intuitively understand the basic electronic properties (band gap, effective mass, etc.) in perovskite semiconductors, and then summarizes recent research progress from two aspects of orbital engineering, thereby providing theoretical guidance for future design of high-performance photovoltaic materials.

## 2. Results and Discussion

### 2.1 Constructing the Orbital-Property Relationship Through Chemical Intuition

It is well known that some basic optoelectronic properties are essential for an efficient photovoltaic device, including suitable band gap, low carrier effective masses, small exciton binding energy, high optical absorption, etc.[36-37] These properties are strongly linked to concepts such as electronegativity, bond length, bond energy, orbital overlap, and band width.[23, 26-27, 38-39] We noticed that Snyder and co-authors have previously used these basic concepts to successfully explain the trends in electronic properties of thermoelectric material such as band gap, carrier effective mass, and band degeneracy and convergence, by using the chemical intuition approach provided by molecular orbital (MO) diagrams, tight binding theory, and a classic understanding of bond strength.[40] Meanwhile, it has been shown that it is very helpful to understand and predict the electronic band structure features without high-level calculations, with the chemical understanding of bonding in solids.[27, 40-41] Therefore, we first take the simplest binary semiconductor MX as an example, and evaluate its band gap and effective mass by employing a simplified model for bonding and band structure evolution near the band gap region. Then we extend the intuitive idea to the family of halide perovskites by choosing specific $AMX_3$ to analyse the trend of electronic properties.

Figure 1a demonstrates the common atomic orbitals involving in reported electronic structures of halide perovskites. Based on MO theory, Figure 1b shows a simplified bonding model to illustrate the evolution of band structure of a polar-covalent semiconductor MX[40-41]. From a chemical point of view, atomic orbitals of element M and element X hybridize with each other, resulting in a bonding states MX and an antibonding state MX* and then forming bands in a solid. In a tight-binding method, a linear combination of such atomic orbitals (LCAO) can be used to describe the electronic structure of a 3D solid. The band gap $E_g$ between the valence and conduction bands in a solid is related to the following bonding parameters in Figure 1b.[40-41]

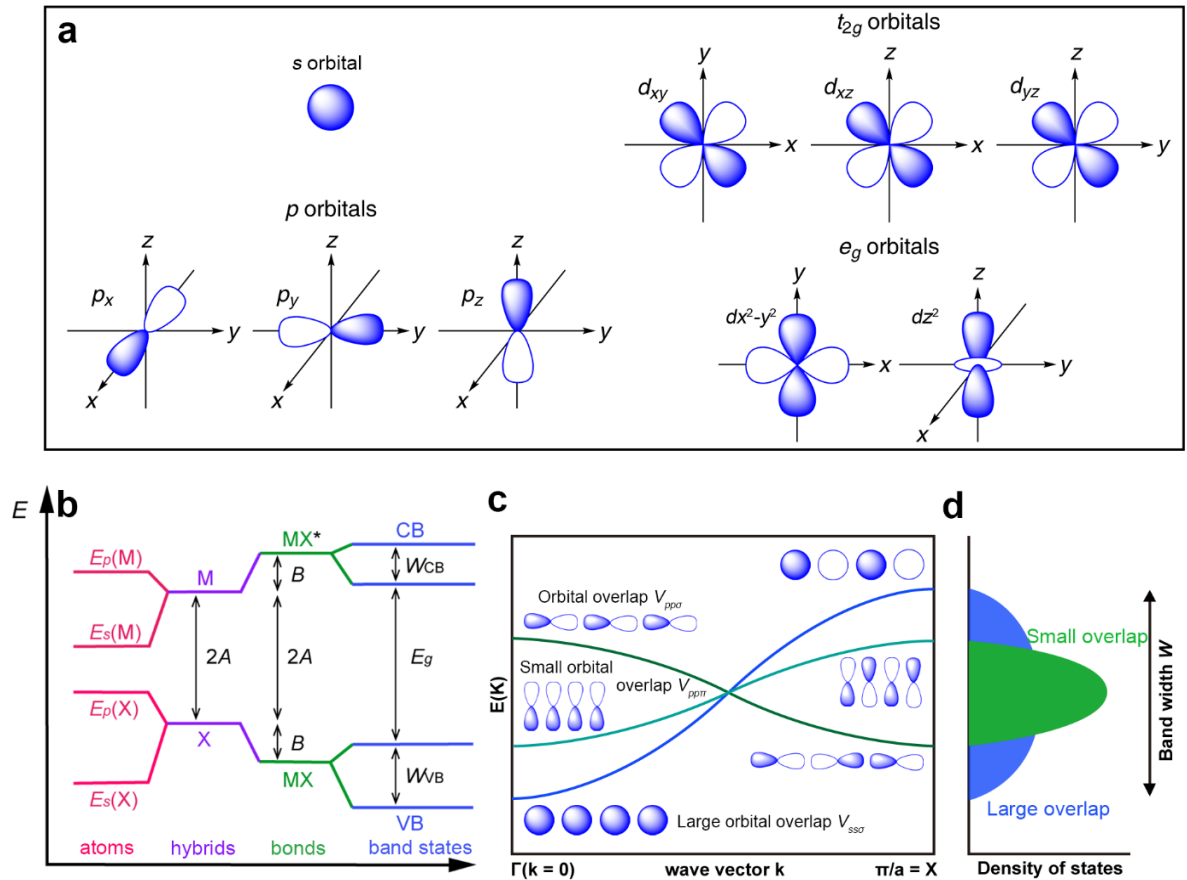

**Figure 1.** (a) The shapes of *s* orbital, *p* orbitals, and *d* orbitals. (b) Transformation of the linear combination of atomic orbital (LCAO), from atomic *s* and *p* orbitals to hybrids, to bond orbitals, and eventually to band states in polar crystal MX.[41] (c) *E* versus *k* dispersion curves in the first Brillouin zone for a periodic array of *s*- and *p*-orbitals with *s* and *p* type interactions. (d) The broader dispersion leads to a lower density of states (DOS) and smaller carrier effective masses in the respective band. (c, d) from ref 40.

(i) 2*A* is the energy difference between the atomic orbitals of M and X, which can be essentially understood as the difference in the ionization energy. Generally, the larger the ionization energy of an element, that is, higher electronegativity, the lower the energy of the atomic orbitals. Previous studies[42-43] have revealed that a larger difference in electronegativity between M and X tend to result in a larger $E_g$ or compounds with larger ionicity tend to have higher $E_g$.

(ii) *B* is the strength of bonding interaction (*i.e.*, bond energy), resulting from the stabilization of the bonding orbital MX and destabilization of the antibonding orbital MX* compared to the respective atomic orbitals. *B* can be expressed as[40]:

$$B = \sqrt{V^2 + A^2} - A, \tag{1}$$

where *V* is the strength of the nearest neighbour coupling. From Figure 1c, it can be seen that a σ-bond between two *s*-orbitals has the largest orbital overlap, and a σ-bond between two *p*-orbitals always has a greater overlap than a π-bond. In all types of interaction, *V* decreases with increasing distance (bond length), *d*, according to

$$V \sim \frac{1}{d^2}. \tag{2}$$

Therefore, increasing *d* lowers the bond strength by decreasing the orbital overlap interaction *V*, which usually leads to a larger $E_g$.

(iii) $W_{VB}$ and $W_{CB}$ are the width of the valence and conduction bands, respectively, which can be simply understood as broadening individual states into a band due to orbital overlap interactions. Generally, band width *W* increases with increasing orbital overlap interaction *V* (small bond length *d*), and increased *W* can reduce the band gap $E_g$ and result in small effective masses, $m^*$, and low density of states (DOS), as shown in Figure 1d.[40]

According to the above discussions, we can conclude that, for polar compounds ($A \neq 0$), a large difference in the energies of the atomic orbitals (large electronegative difference), *A*, will generally decrease the band width (small band dispersion), $W$, since the orbital overlap interactions, $V$, become less effective as $\Delta$ increases. Therefore, ionic materials (large $\Delta$) usually have large band gaps ($E_g$) and narrow bands (small $W$), in which charge carriers are heavy and easily localized (large $m^*$).

**2.1.1 Specific Example Analysis**

The photovoltaic efficiency of the absorber is governed by many factors (i.e., band gap, optical absorption, carrier effective mass, defect tolerance, dielectric constant),[37] among which the band gap is the most critical factor determining the theoretical limit of efficiency according to the Shockley-Queisser limit[44]. Here, we employ the chemically intuitive approach discussed above to understand the trend of band gap in halide perovskites. It is well-established that the M-site cation and X-site anion have the most significant impact on the band gap of $AMX_3$ perovskites, while A-site cation only has an indirect effect on the band edges by distorting the [$MX_6$] inorganic networks due to its highly ionic.[23, 26] Therefore, in the following discussion, we mainly focus on the changing trend of band gap by replacing the elements at the M and X sites.

When M is the alkaline-earth metals ($Ca^{2+}$, $Sr^{2+}$, and $Ba^{2+}$) with the $ns^0$ electron configuration, the VBM consists of $p$ orbital of the X anion, while the CBM is derived from $s$ orbital of the M cation, as shown in Figure 2a.[23] In this case, the band gap is mainly determined by the orbital energy/electronegativity difference (i.e., $\Delta$) between the M cation and the X anion. Figure 2b shows the band gaps of hypothetical orthorhombic phase $CsMI_3$ (M = Ba, Sr, Ca) from DFT calculations with PBE functional.[45] It can be seen that the alkaline-earth metal-based perovskites exhibit relatively large band gaps (> 3 eV) because of the large difference in electronegativity (i.e., large

*A*) between the halogen and alkaline-earth metal atoms (see Figure 2e).[46-47] The observed band gap trend is $E_g(CsBaI_3) > E_g(CsSrI_3) > E_g(CsCaI_3)$ because of an increase in the electronegativity of M cation (see Figure 2e). It is worth noting that the case discussed in Figure 2a is also applicable to explain some oxide and chalcogenide perovskites, in which the VBM consists of *p* orbital of the anion and the CBM is derived from the unoccupied *d* states of the transition metal cation (i.e., M = Ti or Zr).[48-49] They usually also exhibit wide band gaps due to the large difference in electronegativity between the oxygen/sulfur and transition metal atoms.[49]

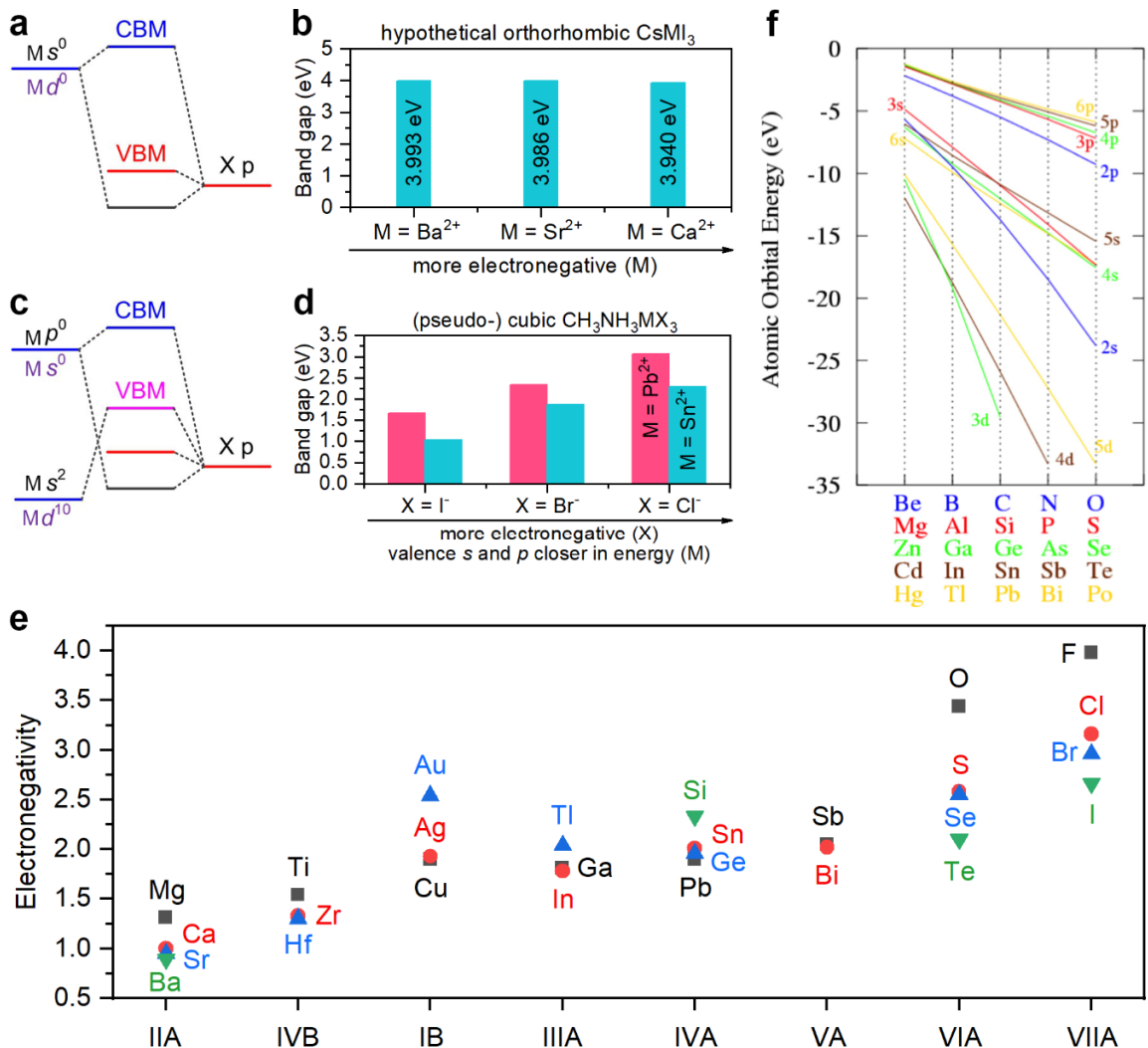

**Figure 2.** Schematic energy diagrams and band gaps for perovskite semiconductors AMX$_3$ (A = Cs, CH$_3$NH$_3$; M = Ca, Sr, Ba, Pb, Sn; X = Cl, Br, I) with VBM being: anion $p$ states (a,b) and antibonding states of cation $s/d$ and anion $p$ orbitals (c,d). (a,c) from ref 23. (d) Electronegativities for representative elements (Pauling units) (f) Chemical trends of atomic energy levels predicted by using the local density approximation (LDA).[50]

When M is the group IVA elements (Ge$^{2+}$, Sn$^{2+}$, and Pb$^{2+}$) with the n$s^2$n$p^0$ electron configuration, the high-energy occupied $s^2$ orbital of M cation can couple with the X $p$ orbital to raise the VBM, therefore narrowing the band gap and increasing the width of valence band ($W_{VB}$), as shown in Figure 2c.[23] In this case, the magnitude of the band gap is not entirely determined by the electronegativity factor. Because the VBM and CBM contain the $s$ and $p$ orbitals of M-site cation, respectively, it is the difference between M-site atomic $s$ and $p$ orbital energies that controls the band gap. Figure 2d shows the computed band gaps of pseudo-cubic phase CH$_3$NH$_3$MX$_3$ (M = Pb and Sn; X = I, Br, Cl) from the GW+SOC method.[51] Due to the small ionic radius of Ge$^{2+}$, Ge$^{2+}$-based compounds usually exhibit a distorted structure,[52] which decreases the orbital overlap interaction $V$, thereby widening the band gap, so they are not included in this discussion. From Figure 2d, it can be seen that the band gap trend at the M site is $E_g$(Pb$^{2+}$) > $E_g$(Sn$^{2+}$) because the valence $s$ and $p$ orbitals of Pb$^{2+}$ differ in energy more (i.e., large $A$) than those of Sn$^{2+}$ (see Figure 2f). Meanwhile, the case discussed in Figure 2c is also applicable to explain the small band gaps in some noble metal cations (i.e., M = Ag or Cu)-containing halide perovskites and chalcopyrites,[49,53] where the $p$-$d$ hybridization between the filled $d$ orbital of metal cation M and the anion X $p$ orbital shifts the VBM levels up.

When varying the X-site element from Cl to Br to I, the observed band gap trend is $E_g(Cl^-) > E_g(Br^-) > E_g(I^-)$ (see Figure 2d).[51] This is also true for other metal cations-based halide perovskites and can be easily understood from the electronegativity of halogen. From Figure 2e, it can be seen that electronegativity for X decreases along the series Cl→Br→I (that is the energy of the atomic orbitals increases in the order Cl $3p$→Br $4p$→I $5p$).[46-47] More electronegativity X-site anions lead to a lower-energy valence band, and therefore a larger band gap.

## 2.2 Identifying the Key Orbital Characters at Band Edges

In various research fields such as photovoltaic,[1] photocatalysis,[54] transparent conducting oxides,[55] thermoelectricity,[56] and heterogeneous catalysis,[57] electronic structure analysis has been widely used as a power tool to understand the basic properties of semiconductor materials at the atomic level. It provides fundamental insights into the band-edge orbital information that are highly correlated with various properties (such as electrical, optical, and defect properties). Recently, Yu et al. also found that the $t_{2g}$-$p$ orbital coupling is responsible for the intrinsic in-plane negative Poisson's ratio in monolayer 1T-MX$_2$ (M = Mo, W, Tc; X = S, Se, Te).[58] In addition, some electronic parameters (i.e, $d$-band center) are also successfully used as descriptors to explain adsorption energy and reaction rate in the field of catalysis.[57, 59] In the field of optoelectronic devices, identifying the key orbital characters at the band edges is very important for screening promising photovoltaic candidates and designing novel functional materials.

### 2.2.1 Screening Promising Photovoltaic Materials

Although lead (Pb) halide perovskites have achieved great success in various fields, the toxicity of Pb is a major factor hindering the commercialization of perovskite optoelectronic devices.[29]

Understanding the intrinsic physical mechanism of the excellent properties of 3D Pb-based halide perovskites is very important to guide the search for Pb-free and stable perovskite alternatives. Electronic structure and Crystal Orbital Hamiltonian Population (COHP) analysis have identified that the superior photovoltaic performance of Pb halide perovskites is mainly attributed to the high perovskite symmetry and the strong antibonding states between Pb 6$s$ and I 5$p$ orbitals at VBM (see Figure 3a-c),[4, 26, 28] which leads to high levels of valence band dispersion (small hole effective masses), ambipolar conductivity and defect tolerance. The unique influence of the existence of n$s^2$ lone-pair cation on appealing properties can be further proved by comparison with the electronic structures of compounds lacking n$s^2$ electron configuration cation (i.e., $Cd^{2+}$ and $Sr^{2+}$). Favini et al. investigated the difference in the band structures of $CsPbBr_3$, $CsCdBr_3$, and $CsSrBr_3$.[28] When $Pb^{2+}$ ($6s^26p^0$) is replaced with $Cd^{2+}$ ($4d^{10}5s^0$) and $Sr^{2+}$ ($5s^0$), the upper valence band becomes much less dispersive, suggesting reduced band width and heavy holes.[28] Therefore, for halide single perovskites, the existence of lone-pair states and the antibonding coupling at VBM is the key band-edge characters, which is critical for favorable optoelectronic properties.

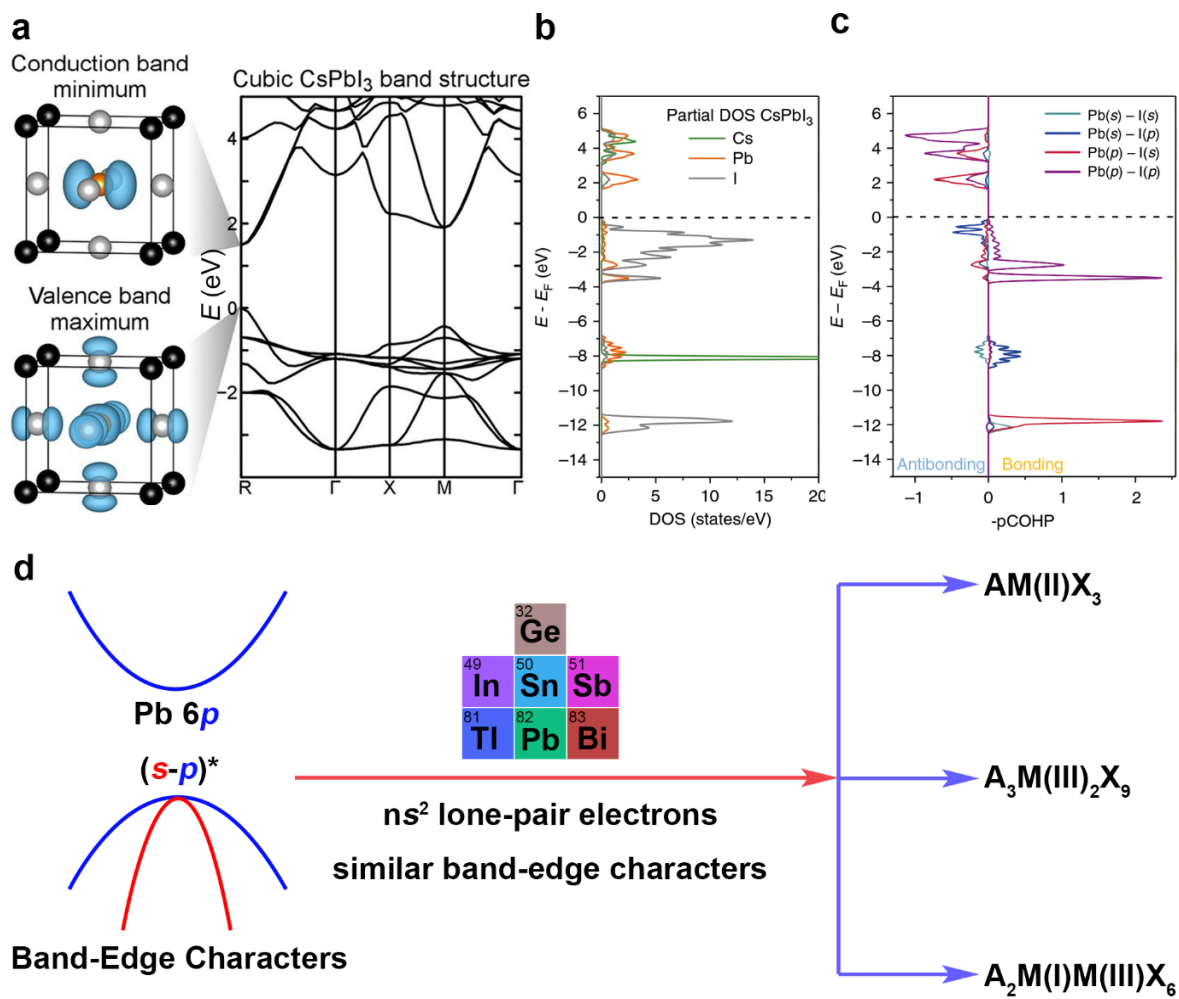

**Figure 3.** (a) Band structure and isosurfaces of the electron densities, (b) projected density of states (DOS), (c) orbital-resolved crystal orbital Hamiltonian population (COHP) of CsPbI$_3$. (a-c) from ref 60. (d) Illustration of methodology for theoretically identifying promising perovskite candidates from key band-edge features.

Guided by the key features of electronic structure, a series of n$s^2$-containing (M = Sn$^{2+}$, Ge$^{2+}$, Sb$^{3+}$, Bi$^{3+}$) halide perovskites have been widely explored as potential optoelectronic materials,[3, 20, 29] as shown in Figure 3d. In these compounds, although similar band-edge characters are retained, new instability problems and reduced structural dimensionality are introduced. For example, Sn$^{2+}$ and $Ge^{2+}$-based perovskites suffer from serious instability issues (i.e., against oxidation to $Sn^{4+}$ and $Ge^{4+}$),[29] leading to deterioration of photovoltaic performance. To maintain the neutrality of the charge, trivalent cations such as $Bi^{3+}$ and $Sb^{3+}$-based perovskites $A_3M_2X_9$ adopt low-dimensional structures, resulting in the weak antibonding coupling at VBM and thereby undesired photovoltaic properties such as wide band gap, large hole effective mass, and defect intolerance.[3] In addition to those materials with the lone-pair $ns^2$ cation, the search for new compounds only containing strong antibonding coupling at VBM is also exploited. For example, Yin et al. have reported that $IrSb_3$ with a skutterudite-structure possesses similar electronic properties as $CH_3NH_3PbI_3$, which are derived from the *p-p\** antibonding coupling at VBM.[25] Unfortunately, the low-temperature synthesis of high-quality thin film of $IrSb_3$ is challenging, and no related solar cells have been reported.

Based on the above discussions, a promising high-performance Pb-free absorber should include a relatively stable lone-pair cation (i.e., $Bi^{3+}/Sb^{3+}$) in high-symmetry coordination and a high degree of connectivity between atoms.[28] To date, one of the most promising approaches is to substitute the two Pb(II) in $APb(II)X_3$ into a pair of M(I)/M(III) cations to form 3D halide double perovskites $A_2M(I)M(III)X_6$ (see Figure 4a and 4b).[22] Slavney et al. first experimentally synthesized the cubic $Fm\overline{3}m$ double perovskite $Cs_2AgBiBr_6$, which exhibits wide indirect band gap of 1.95 eV.[15] Subsequently, Volonakis et al. synthesized a new direct band gap double perovskite $Cs_2InAgCl_6$ with anomalously large optical band gap of 3.3 eV.[18] It can be seen that despite the combination of a 3D crystal structure and a lone pair cation (e.g., $Bi^{3+}$ or $In^+$), the synthesized double perovskites do not exhibit desired optoelectronic properties. Therefore, some theoretical studies[21-22] have been carried out to reveal that in order to achieve superior performance of $A_2M(I)M(III)X_6$ double perovskites, the key band-edge features are that the frontier atomic orbitals between the M(I) and M(III) cations should be electronically matched, and these band-edge orbitals should have a high dimensionality of connectivity at the VBM and CBM.

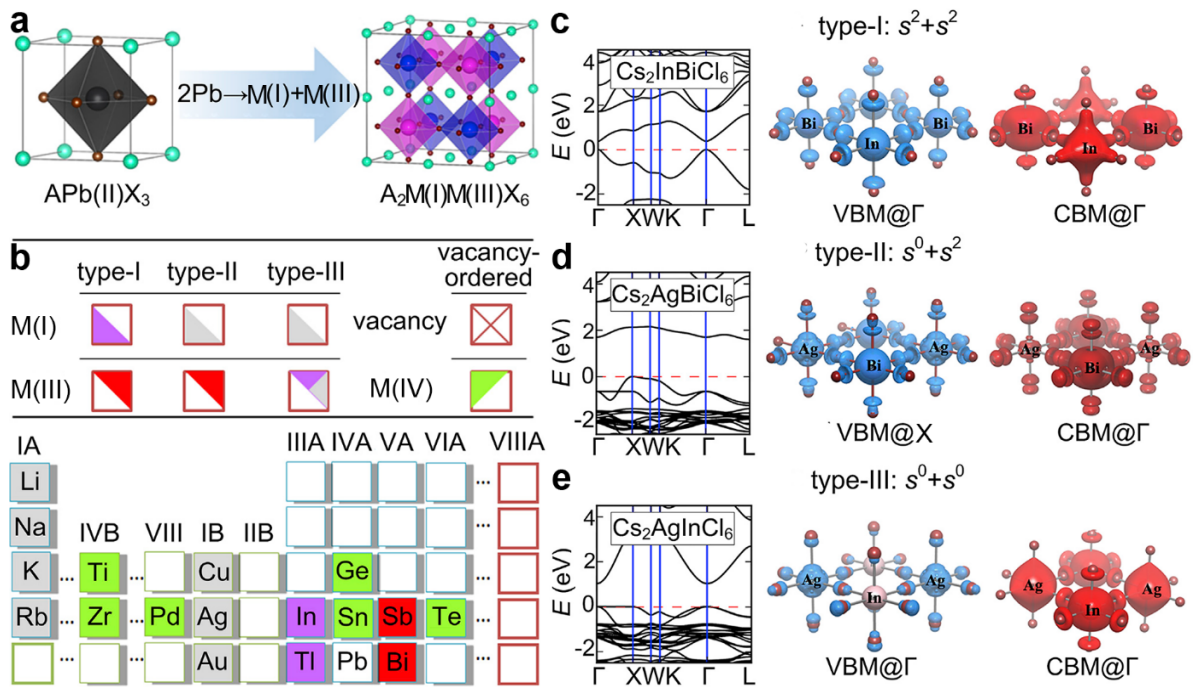

**Figure 4.** (a) Schematic illustration of the cation-transmutation from single perovskites to double perovskites. (b) Classification of halide double perovskites and the possible combination space for the M-site cations. Band structures (left) and electron densities of VBM (middle) and CBM (right) of representative halide double perovskites: (c) type I: $Cs_2InBiCl_6$, (d) type II: $Cs_2AgBiCl_6$, (e) type III: $Cs_2AgInCl_6$. (a-f) from ref 22.

Considering the importance of lone-pair states in single perovskites, in order to fully understand the diverse optoelectronic properties of halide double perovskites, it is reasonable to classify them based on whether M(I) or M(II) cation contains lone-pair states, into three types (see Figure 4c-e):

(i) both M(I) and M(III) cations contain lone-pair electrons (type I: $s^2+s^2$); (ii) only one of M(I) and M(III) cations has lone-pair electrons (type II: $s^0+s^2$); (iii) neither M(I) nor M(III) cation includes lone-pair electrons (type III: $s^0+s^0$).[22] Figure 4c shows the case of the hypothetical compound $Cs_2InBiCl_6$ (type I: $s^2+s^2$). The combination of In(I) and Bi(III) cations matches the frontier orbitals because they have the same fully occupied outmost $s$ shells similar to Pb(II). As expected, the band structure (see Figure 4c) of $Cs_2InBiCl_6$ is quite similar to those of $APbX_3$. Specifically, the band gap is direct at the Γ point, and highly dispersive band edges (i.e., K-Γ-L direction) are also observed. The small carrier effective masses and large band widths can be well explained from the uninterrupted In-Cl-Bi connectivity at the atomic level. For example, the CBM derives from the antibonding states of Bi $6p$/In $5p$-Cl $3p$ orbitals, and the VBM consists of the antibonding states of In $5s$/Bi $6s$-Cl $3p$ orbitals. Meng et al. also revealed that $s^2+s^2$ type $A_2M(I)M(III)X_6$ (M(I) = $In^+$, $Tl^+$; M(III) = $Sb^{3+}$, $Bi^{3+}$) double perovskites show favorable optical absorption suitable for thin-film solar cell applications.[61] Unfortunately, $Tl^+$ is more toxic than $Pb^{2+}$ and $In^+$-based double perovskites tend to be unstable against oxidation or form mixed-valence compounds with distorted and complex crystal structures.[16, 62-64] At present, $(CH_3NH_3)_2TlBiBr_6$ is the only one synthesized example of $s^2+s^2$ type,[65] and the corresponding solar cells have not been reported.

When replacing half of the lone-pair cations with $d^{10}$ cations such as $Ag^+$, as shown in Figure 4d, $Cs_2AgBiCl_6$, as a typical example of $s^0+s^2$ type, exhibits an indirect band gap with the CBM at the L point and the VBM at X point. It has been established that the VBM of $Cs_2AgBiCl_6$ is composed of A$d$ $4d$/Bi $6s$-Cl $3p$ antibonding states, and the CBM mainly consists of Bi $6p$ states. Savory et al. first revealed that the indirect characteristics and less band-edge dispersion (i.e., X-W direction) in $Cs_2AgBiBr_6$ are attributed to the mismatch in angular momentum of the frontier atomic orbitals (i.e., Ag 4$d$ and Bi 6$s$).[21] For the $s^0+s^0$ type, the representative material $Cs_2AgInCl_6$ shows a direct band gap of 2.1 eV at the Γ point (see Figure 4e). In the case, the CBM consists of In 5$s$/Ag 5$s$ and Cl 3$p$ orbitals, while the VBM is derived from Ag 4$d$ and Cl 3$p$ orbitals. Zhao et al. demonstrated that the atomic orbitals near the valence band edge cannot connect 3-dimensionally because the [$AgCl_6$] octahedra are isolated by the adjacent [$InCl_6$] octahedra,[22] leading to 0D-electronic features such as flat band along the Γ-X direction and very large hole effective mass. In addition, both theoretical and experimental studies have confirmed that $Cs_2InAgCl_6$ exhibits inversion symmetry-induced parity-forbidden transitions between band edges,[18, 61] which is undesirable for photovoltaic applications.

The above-identified key electronic features of halide perovskites may also have important guiding significance for the design of promising chalcogenide single (double) perovskites and emerging layered halide double perovskites. At present, the realization of Pb-free and stable photovoltaic materials with excellent performance comparable to 3D $APbI_3$ still requires extensive theoretical and experimental investigation.

**2.2.2 Designing Novel Functional Materials**

The reasonable manipulation of band characters near the Fermi-level provides an effective method to achieve novel properties in functional materials.[66] A typical example is to utilize the different running way of $s$- and $p$-orbital-derived bands (see Figure 5a) to adjust the direct-indirect band gap transition in semiconductors.[17] As shown in Figure 5b, Tran et al. proposed a design strategy to tune the convergence of direct and indirect band gaps based on chemically adjusting the $s$- and $p$- orbital character of the CBM,[17] which can be realized by adjusting the electron filling of the M(III) cation from $s^0$ to $s^2$ in $A_2M(I)M(III)X_6$ (M(I) = $Cu^+$, $Ag^+$, $Au^+$) halide double perovskites. Specifically, the VBM is derived from the fully occupied $d$ orbitals of M(I) cation, while the CBM depends on choosing the appropriate M(III) cation that can contains the $s/p$ states. In this case, it is then possible to adjust the CBM from $s$-orbital-derived to $p$-orbital-derived while leaving the remainder of the band structure essentially intact. To confirm the viability of this design strategy, Tran et al. successfully synthesize two Pb-free double perovskites $Cs_2AgSbCl_6$, with an indirect bandgap, and $Cs_2AgInCl_6$, with a direct bandgap, and demonstrating a crossover from indirect to direct optical absorption in a solid solution between the two at 40%:60% Sb:In (see Figure 5b).[17] It should be pointed out that some other theoretical studies believe that whether there are M(III) $s$ states at the VBM is also very critical for the direct/indirect band gap character of $Cs_2AgM(III)X_6$ (M(III) = $In^{3+}$, $Bi^{3+}$, $Sb^{3+}$).[21]

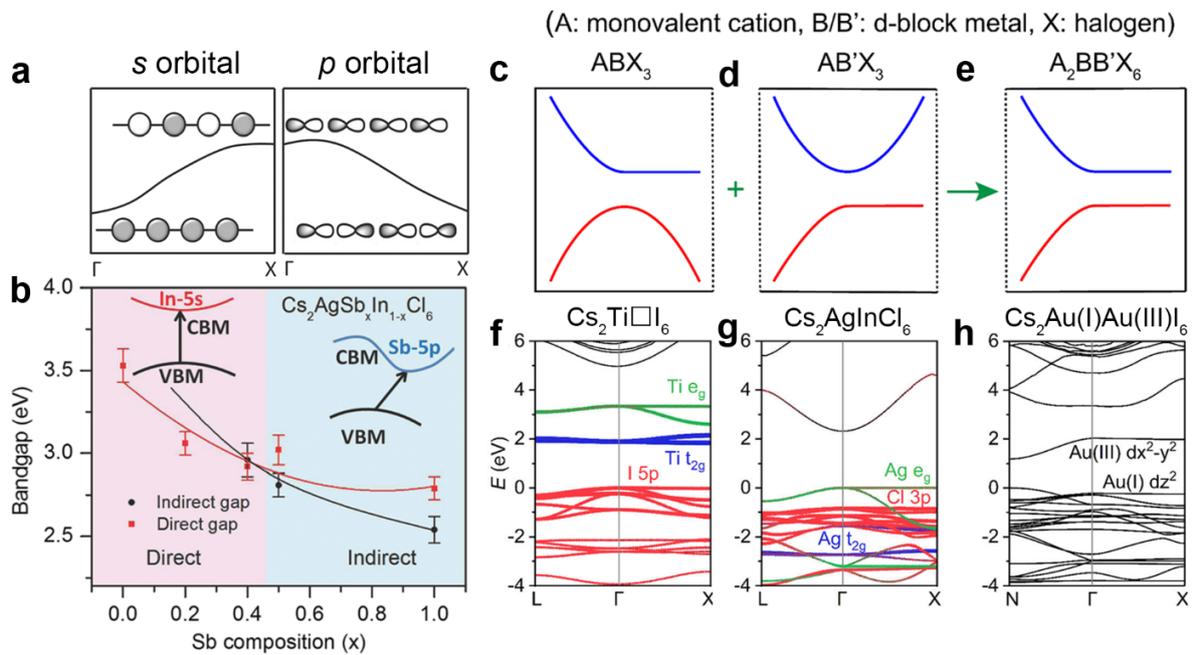

**Figure 5.** (a) Band structures for $s$- and $p$-orbital derived bands. (b) Phase diagram for $Cs_2AgSb_xIn_{1-x}Cl_6$ showing the bandgap trend as a function of Sb composition and a crossover from indirect to direct optical absorption in a solid solution at $x = 0.4$. (a,b) from ref 17. (c-e) Schematics of orbital engineering for *d*-block metal cation-based halide perovskites. Panels c and d show the B (e.g., empty d orbital)/B'(e.g., fully filled d orbital) *d*-X *p* hybridization creating the 2D character of the conduction/valence bands. Panel e shows that 2D band edges are achieved by combining the two different *d*-block metals. To validate the design ideas, (f-h) shows the calculated band structures of three specific materials such as $Cs_2TiI_6$, $Cs_2AgInCl_6$, and $Cs_2Au(I)Au(III)I_6$ corresponding to the three cases in panels c-e, respectively. (c-h) from ref 19.

Another typical example is to control the flat band near the band edges to achieve novel properties,[67-68] which has been successfully applied to obtain high power factor in the field of thermoelectricity.[69] Recently, Tang et al. proposed a design strategy to construct 2D electronic structures in 3D halide perovskites by reasonably controlling the flat band character at the band edges.[19] As shown in Figure 5c-e, the occurrence of the non-dispersive band at the conduction/valence band edge will depend on the degree of *d* orbital filling of the metal cation because the *d-p* hybridization tends to create low-dimensional electronic structures. If the cation *d* orbital is fully empty (fully filled), the metal-halide orbital interactions will result in the presence of flat band at the conduction (valence) band edge (see Figure 5c,d). However, for the partially filled cases, the situation is complicated and depends on the actual material systems. Therefore, it is possible to achieve flat bands both at the conduction and valence band edges simultaneously by combining two metal cations with different *d*-electron configurations (see Figure 5e).

To verify the feasibility of this strategy, Tang et al. employed specific materials as a case study to analyse their electronic structures.[19] As the representative compound of $d^0$ type, vacancy-ordered double perovskite $Cs_2TiI_6$ crystallizes in the cubic crystal structure with space group $Fm\bar{3}m$,[70] and its band structure is shown in Figure 5f. As expected, the lower conduction band of $Cs_2TiI_6$ exhibits non-dispersion along Γ-X direction, originating from the hybridization of I $5p$ states and Ti $t_{2g}$ states. In addition, it has also been reported that the partially filled Fe $e_g$ states can also form flat conduction bands at the band edge along Γ-X direction in the class of $Fe_2YZ$ (Y = Ti, Zr, Hf; Z = Sn, Ge, Si) full Heusler compounds.[69] $Cs_2AgInCl_6$ is chosen as the representative perovskite of $d^{10}$ type,[18] and the calculated band structure is shown in Figure 5g. Indeed, the flat band between Γ and X occurs at the valence band edge, mainly resulting from the hybridization of Cl $3p$ states and Ag $e_g$ states. Finally, to combine two $d$-block metal elements with different electron filling in the same perovskite structure while satisfying the charge neutrality, the experimentally synthesized mixed-valence double perovskite $Cs_2Au(I)Au(III)I_6$ is selected as an example.[71] As shown in Figure 5h, it can be seen that the flat conduction band and valence band at the band edges are achieved simultaneously along the Γ-X direction, which results in the 2D electronic properties in a pure 3D halide perovskite. They further revealed that the weak coupling between I $5p$ states and Au(I) $5d_{z^2}$ states is responsible for the flat valence band, and the flat conduction band originates from the hybridization of I $5p_{x,y}$ states and $5d_{x^2-y^2}$ states of Au(III) and Au(I), leading to a 2D wave function confined within the equatorial (001) plane.

## 2.3 Manipulating the Orbital Components of Band Edges

To realize excellent performance of halide perovskites, a series of methods and strategies, such as compositional engineering,[34] dimensionality engineering,[10] strain engineering,[35] orbital-splitting approach,[72] etc. have been reported to engineer their band-edge electronic features. In this section, we will use the above established orbital-property relationship to understand how these methods modify the band-edge orbital components and band-edge positions, thereby regulating the optoelectronics properties of halide perovskites.

**2.3.1 Compositional Engineering**

Replacing the elements at the A-site, M-site or X-site is the most straightforward way to change the band-edge orbital components.[60] The influence of $M^{2+}$ and $X^-$ ions on the band gap has been discussed above, and here we mainly focus on the changing trend of band-edge positions reported by Tao et al. As shown in Figure 6a and 6d,[60] when going from I to Br to Cl, the increase in the band gap is dominated by the downward shift of the VBM, although the slight upward shift of the CBM is also observed. The increase in electronegativity from I to Br to Cl (see Figure 2e) leads to a significant downward shift of the $X^-$ $p$ atomic level, which is the dominating factor in determining the position of the VBM. Due to the confinement effect (i.e., the Pb-X distance from I to Br to Cl decreases), the energy of the Pb 6$p$ and 6$s$ atomic levels shift upward, which affects the position of the CBM and the VBM, respectively.

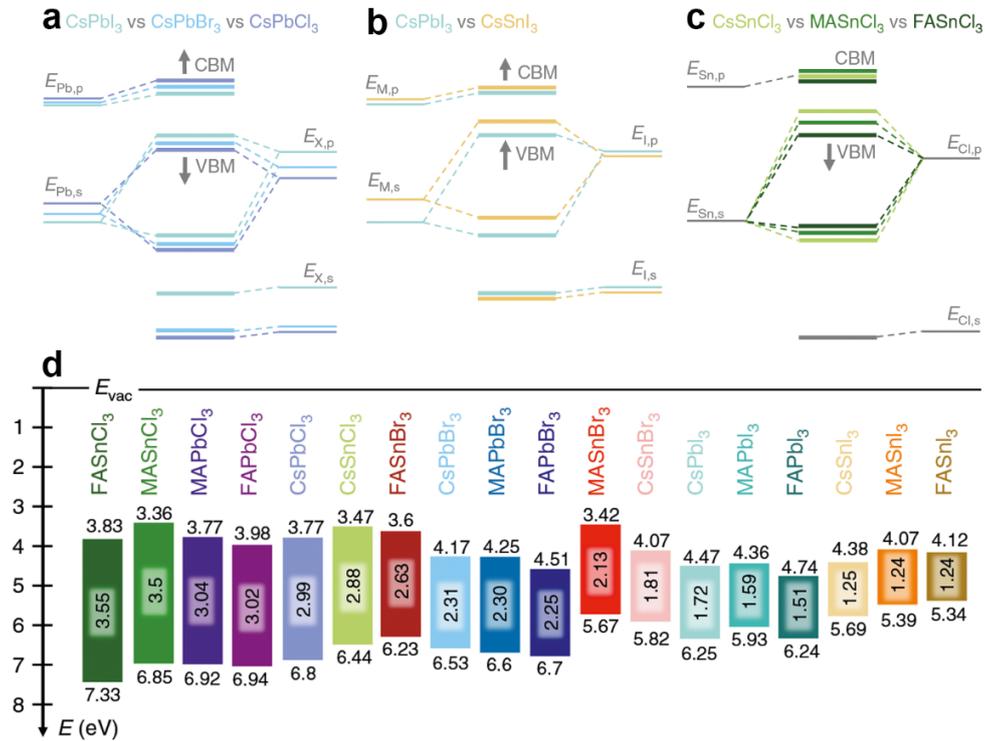

**Figure 6.** Schematic energy levels in AMX$_3$ perovskites. (a, b) Represent trends in changing the halide anions and the metal cations, respectively, as identified from the tight-binding analysis. (c) is an intuitive illustration of energy-level changes based on structural distortions in tin-based perovskites. Arrows indicate the shift in energy levels upon atom or small molecule substitution. (d) Schematic energy level diagram of the 18 metal halide perovskites. (a-d) from ref 60.

When Pb is replaced with Sn, it can be observed that both the VBM and the CBM move upward, as depicted in Figure 6b and 6d.[60] The movement of the CBM and the VBM are determined by the upward shift of the M$^{2+}$ $p$ and $s$ atomic levels, respectively, because Sn has a smaller electronegativity than Pb (see Figure 2e). Moreover, it is worth noting that the upward shift of the $s$ level is larger than that of the $p$ level, indicating the VBM shifts upward more than the CBM. The smaller splitting between $s$ and $p$ states explains the generally reduced band gaps of Sn$^{2+}$-based compounds compared to their Pb$^{2+}$ counterparts.

Unlike M-site cation and X-site anion, substituting the A-site cation only indirectly influences the electronic structure because it does not directly contribute to the band edges. This indirect effect includes changing the volume of the $AMX_3$ lattice or distorting the $[MX_6]$ inorganic framework. The influence of these two factors on the band edges will be discussed in detail in the following section. Here just simply take $ASnCl_3$ series as an example (see Figure 6c and 6d),[60] when going from Cs to MA to FA, a large downward shift of the VBM is the dominating factor in determining the increase of the band gap. This is because the increased lattice distortion (tolerance factor, $t_{CsSnCl_3} < t_{MASnCl_3} < t_{FASnCl_3}$) weakens the anti-bonding coupling at the VBM.

**2.3.2 Dimensionality Engineering**

Keeping the elements fixed and changing the dimensionality and connectivity of $[MX_6]$ octahedral framework also has an important impact on the orbital components near the band edges.[10-14] As shown in Figure 7a, Xiao et al. used $CsPbI_3$, $Cs_2PbI_4$, $Cs_3PbI_5$, and $Cs_4PbI_6$ as representatives of 3D, 2D, 1D and 0D perovskites, respectively, to study the effect of structural factors on the optoelectronic properties.[11] The band structures (see Figure 7b) reveal that as the dimensionality of the octahedral network decreases, the band gap gradually increases, the bands near the Fermi level become less dispersion, the band width becomes narrower, and the carrier effective masses become larger. This can be well explained by the reduced connectivity of atomic orbitals that contribute to the band edges (i.e., reduced electronic dimensionality).

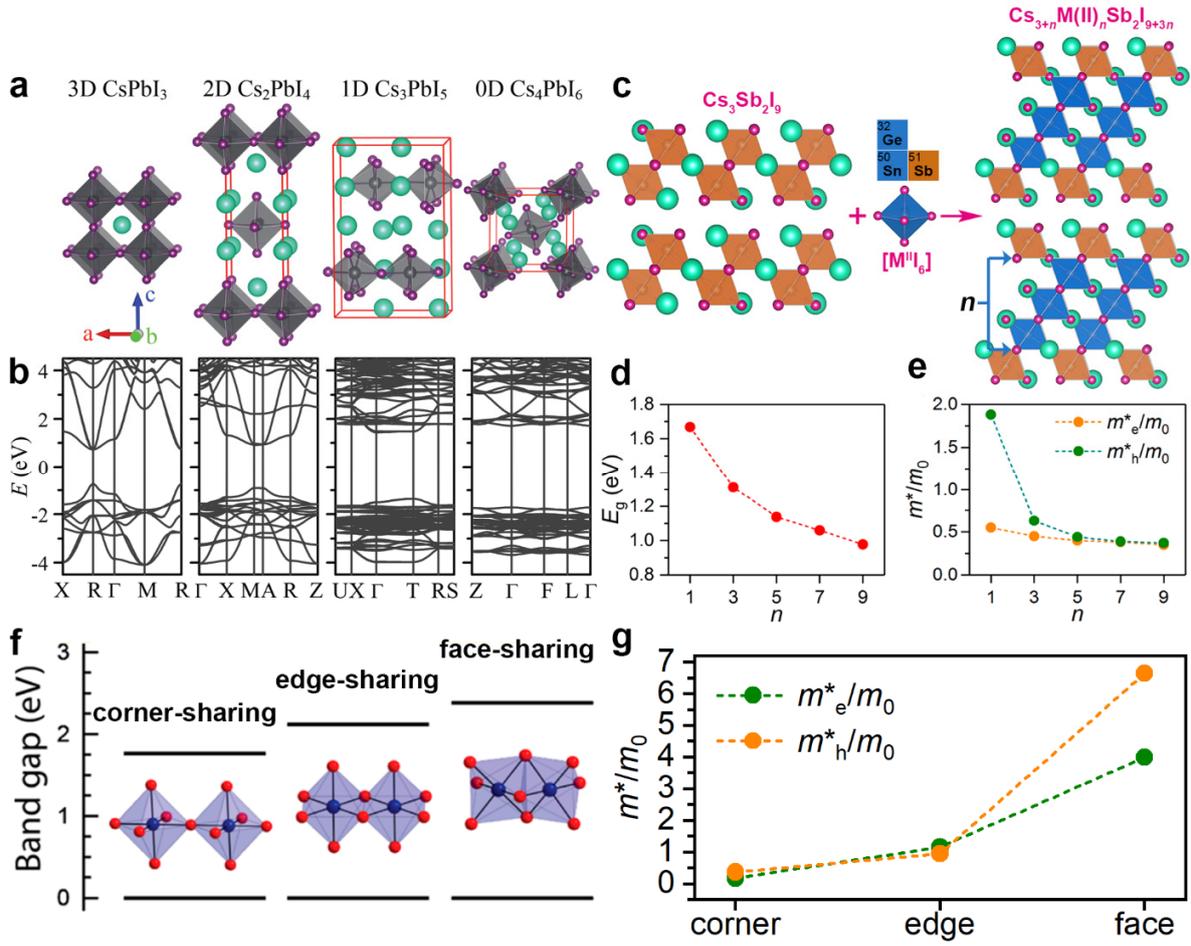

**Figure 7.** (a) Hypothetical model crystal structures and (b) calculated band structures of $CsPbI_3$, $Cs_2PbI_4$, $Cs_3PbI_5$, and $Cs_4PbI_6$. (a-b) from ref 11. (c) Illustration of the general design principle for the layered halide double perovskites $Cs_{3+n}M(II)_nSb_2X_{9+3n}$ (M = Sn, Ge) by inserting $[M^{II}X_6]$ octahedral layers into the $[Sb_2X_9]$ bilayers. Left and right panels show the crystal structures of $Cs_3Sb_2I_9$ and hypothetical $Cs_{3+n}M(II)_nSb_2I_{9+3n}$ with $n$ = 3, respectively. (d) Calculated band gaps and (e) carrier effective masses of layered double perovskite $Cs_{3+n}Sn_nSb_2I_{9+3n}$ with $n$ = 1, 3, 5, 7, and 9. (c-e) from ref 73. (f) Approximate band gaps of theoretical model structures with different connectivity.[14] (g) Carrier effective masses of $GAPbI_3$ (GA = $C(NH_2)_3$) with hypothetical 3D corner-sharing, experimental 1D edge-sharing, and hypothetical 1D face-sharing structures.[12]

In particular, for layered perovskites, the photovoltaic performance strongly depends on the layer number $n$ of [MX$_6$] octahedra. As shown in Figure 7c-e, Tang et al. calculated the layer-dependent optoelectronic properties of hypothetical layered halide double perovskites Cs$_{3+n}$Sn$_n$Sb$_2$X$_{9+3n}$.[73] The band gap and carrier effective masses decreases significantly as $n$ increases from 1 to 9, originating from the enhanced band-edge orbital overlap between Sb and Sn/5$s$–I/5$p$ (Sb/5$p$–Sn/5$p$) pushing up the VBM due to the increased [SnI$_6$] octahedral layers.

For low-dimensional perovskites, the octahedral connectivity may appear edge-sharing and face-sharing instead of corner-sharing. As shown in Figure 7f and 7g, Kamminga et al.,[14] Deng et al.,[13] and Tang et al.[19] investigated the effect of the connectivity of [MX$_6$] octahedra on the optoelectronic properties. It can be clearly seen that increasing the connectivity from corner-, via edge-, to face-sharing causes a significant increase in the band gap and carrier effective masses. This is because that an increase in the average octahedral M-X distances and M-X-M angles from corner-, to edge-, to face-sharing structures significantly weak the band-edge orbital hybridization.

### 2.3.3 Strain Engineering

Strain effects such as interface lattice mismatches, local inhomogeneous composition distribution, high pressure, etc. have been widely reported to have an important impact on the photovoltaic performance and stability of perovskite solar cells.[26, 35, 74] It can be simply decomposed into two factors: lattice variation and structural distortion (i.e., tilting and rotation of the [MX$_6$] octahedra). Figure 8a displays a schematic energy level diagram showing how each kind of deformation affects the valence and conduction bands.[75] It can be seen that, compared to the CBM composed of metal $p$ and halide $p$ orbitals with less antibonding and more nonbonding character, the VBM, derived from the antibonding states of metal *s* and halogen *p* orbitals, is more sensitive to lattice changes and structural distortion. Therefore, strain engineering is an effective approach to tune the optoelectronic properties by changing the shape and position of the VBM.

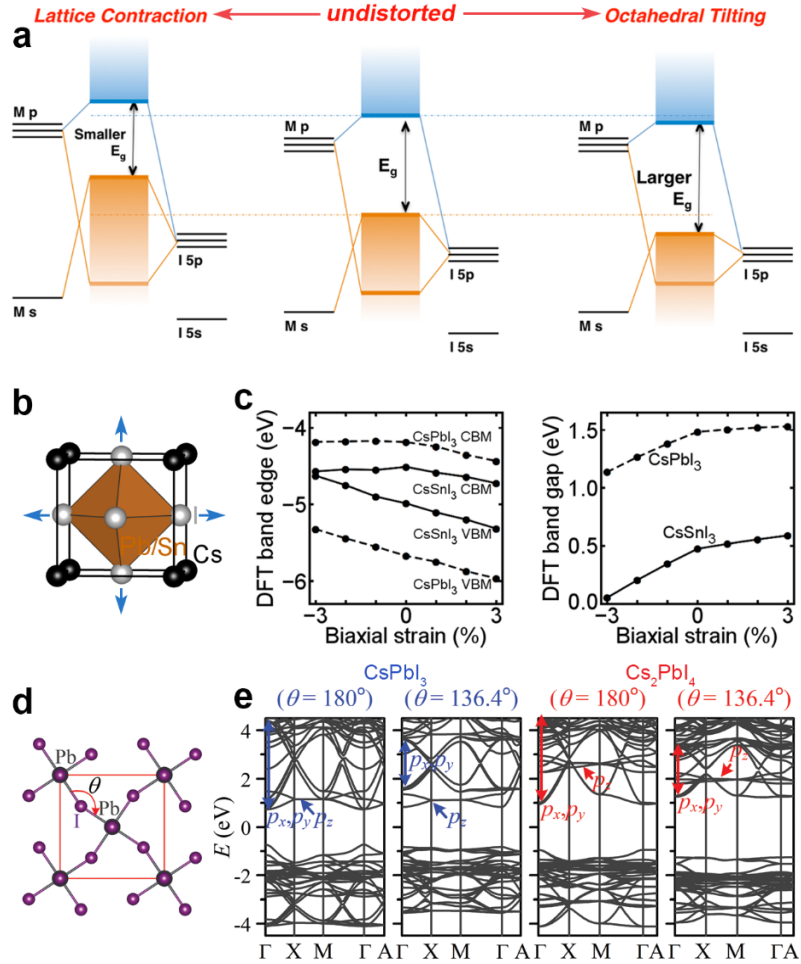

**Figure 8.** (a) The schematic energy level diagram of undistorted cubic lattice (center panel), lattice contraction (left panel) and octahedral tilting (right panel).[75] (b) Cubic crystal structure of CsMI$_3$ (M = Pb and Sn). (c) Absolute band-edge energies (relative to the vacuum level) (left panel) and band gaps under biaxial strain are shown for CsSnI$_3$ (solid lines) and CsPbI$_3$ (dashed lines) (right panel). (b,c) from ref 76. (d) Schematic of Pb-I-Pb bond distortion in the *ab* plane of $\sqrt{2} \times \sqrt{2}$-

cells of CsPbI$_3$ and Cs$_2$PbI$_4$ perovskites. (e) Calculated band structures for ideal undistorted ($\theta$ = 180º) and distorted ($\theta$ = 136.4º) CsPbI$_3$ and Cs$_2$PbI$_4$ perovskites. (d,e) from ref 11.

Grote et al. studied the effect of biaxial strain on the band-edge energies and band gaps of cubic CsMI$_3$ (M = Sn and Pb),[76] as shown in Figure 8b and 8c. As expected, the CBM is less affected by strain because changes in bond lengths have little bearing on nonbonding orbitals. And the band gaps of both CsSnI$_3$ and CsPbI$_3$ increase monotonically from compression to tension strain due to the lowering of the VBM. This can be well understood from previous analysis that the lattice expansion (contraction) weakens (increases) the M-X orbital coupling and then lowers (raises) the energy of the VBM (see Figure 8a).

Xiao et al. studied the effect of structural distortion on the electronic properties of ideal 3D CsPbI$_3$ and 2D Cs$_2$PbI$_4$ perovskites,[11] as depicted in Figure 8d and 8e. It can be seen that the structural distortion significantly alters band-edge orbitals (i.e., Pb $p_x$/$p_y$/$p_z$) and band dispersion (widths). Specifically, as the in-plane Pb-I-Pb bond angle $\theta$ decreases from 180º to 136.4º, the band gap of 3D CsPbI$_3$ increases slightly from 1.48 to 1.66 eV, while that of 2D Cs$_2$PbI$_4$ increases significantly from 1.90 to 2.51 eV. The reason for the increase (reduce) in the band gap (band widths) is that the octahedral rotation decreases the M-X orbital coupling and lowers the energy of the VBM (see Figure 8a).

### 2.3.4 Orbital-Splitting Approach

Orbital-splitting induced by spin-orbit coupling (SOC) and symmetry reduction has an important influence on the band-edge orbital components,[72, 77] which significantly affects the optoelectronic properties of perovskite semiconductors. Even et al. first revealed that SOC effect significantly reduces the band gap of Pb halide perovskites by inducing a large splitting of the first degenerated conduction levels (see Figure 9a and 9b).[77] While the valence bands are nearly unaffected by the SOC effect. Therefore, it is necessary to consider the SOC effect in the electronic calculation when predicting the promising photovoltaic candidates containing heavy elements (i.e., Pb and Bi).

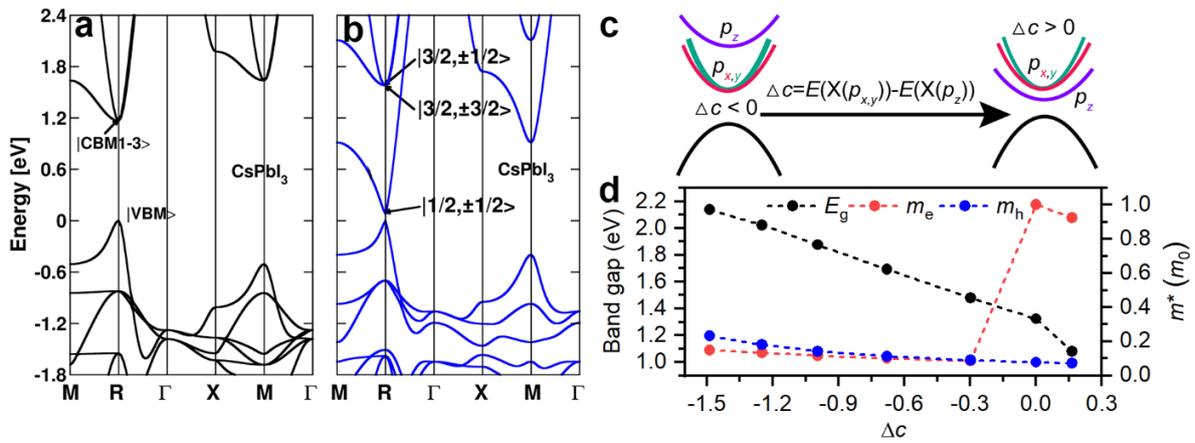

**Figure 9.** Electronic band structures of $CsPbI_3$ in the cubic phase without (a) and with (b) the spin-orbit coupling (SOC) interaction. (a,b) from ref 77. (c) Schematic diagram of orbital engineering by manipulating the relative energies of $p_x$, $p_y$ and $p_z$ orbitals at a specific high symmetry point (e.g., X). Nondegenerate band $X(p_z)$ and doubly degenerate band $X(p_{x,y})$ are mainly composed of $p_z$ and $p_{x,y}$ orbitals from metal cations, respectively. $\Delta c$ is the splitting energy between $p_{x,y}$ and $p_z$ orbitals at the X point. (d) Band gap and carrier effective masses as a function of splitting energy $\Delta c$ in the representative Ruddlesden-Popper (RP) phase perovskite $Cs_2GeI_2Cl_2$. (c,d) from ref 33.

The reduction in crystal symmetry induced by dimensionality reduction (see Figure 7a) or structural distortion (see Figure 8d) is very common in halide perovskites. For examples, in parent $CsMCl_3/CsMI_3$ cubic bulk perovskites ($Pm\overline{3}m$), the conduction band edge consists of three-fold degenerated *p* orbitals at R point (1/2, 1/2, 1/2) of the cubic Brillouin zone (see Figure 3a).[60] For the Cs$_2$MI$_2$Cl$_2$ layered perovskites (*I4/mmm*), the conduction band edge is still derived from *p*-states of the M-site metal atoms, but it is located at X point (1/2, 1/2, 0) of the tetragonal centred Brillouin zone and the $p_z$ orbital splits at higher energy than the $p_x$ and $p_y$ degenerated states (see Figure 9c).[33] Taking Cs$_2$GeI$_2$Cl$_2$ as an example, Tang et al. recently proposed an orbital-splitting approach to explore the orbital-property relationship in halide perovskites.[33] As shown in Figure 9c and 9d, they defined the energy difference between the $p_{x,y}$ orbital-derived bands and $p_z$ orbital-derived band as the splitting energy, namely, $\Delta c = E(X(p_{x,y})) - E(X(p_z))$, and studied the evolution of optoelectronic properties such as band gap ($E_g$) and in-plane carrier effective masses ($m^*$) with the change of $\Delta c$. It can be seen that $E_g$ and $m^*$ decreases monotonically with the decreasing $|\Delta c|$ value, originated from the enhanced orbital hybridization in the in-plane direction. However, when $\Delta c$ becomes positive values, meaning the energy levels of $p_{x,y}$ and $p_z$ orbitals are exchanged and $p_z$ orbital-derived band becomes the new CBM at X point, the band gap shows a large drop because the energy level of $p_z$ orbital is more sensitive to the interlayer hybridization than that of $p_{x,y}$ orbitals. And the band width from Ge $p_z$-derived band is significantly smaller than that of Ge $p_{x,y}$-derived bands can explained the increases of $m_e$.

## 3. Conclusion

In summary, this Perspective has developed the basic concepts such as atomic orbital energy and electronegativity to help intuitively understand the trends in electronic properties of perovskite semiconductors, and then summarized the key band-edge orbital character for designing superior halide single and double perovskites, finally reviewed the changing trends in electronic properties by manipulating the orbital components of band edges. Some methods and strategies presented here can also be extended to understand the optoelectronic properties of chalcogenide perovskites and nitride perovskites. Based on the identified key electronic features near the band edges, it is still challenging for Pb-free and stable perovskite photovoltaic candidates to achieve performance comparable to Pb halide perovskites. We believe that this Perspective can provide atomic-level insights for the future theoretical design of superior photovoltaic materials and experimental optimization of device performance.

## AUTHOR INFORMATION

**Corresponding Author**
*E-mail: hongjw@bit.edu.cn.
**Notes**
The authors declare no competing financial interests.

## ACKNOWLEDGMENT

This work is supported by the National Science Foundation of China (Grant No. 11572040), Beijing Natural Science Foundation (Grant No. Z190011) and Technological Innovation Project of Beijing Institute of Technology.